\documentclass[aps,prl,reprint,twocolumn,superscriptaddress,amsmath,amssymb,amsfonts]{revtex4-1}

\AtBeginDocument{\usepackage{booktabs}}
\makeatletter
\g@addto@macro\bfseries{\boldmath}
\makeatother

\usepackage[varg]{txfonts}
\usepackage[T1]{fontenc}
\usepackage[utf8]{inputenc}

\usepackage{color}
\usepackage{hyperref}
\definecolor{cL}{RGB}{32,145,140}
\hypersetup{colorlinks=true,linkcolor=cL,citecolor=cL,urlcolor=cL} 

\usepackage{float}
\usepackage{multirow}
\usepackage{amsmath}
\usepackage{graphicx}
\usepackage[percent]{overpic}
\usepackage{mathrsfs}
\usepackage{bm}
\usepackage{braket}
\usepackage[nolist,nohyperlinks]{acronym}
\usepackage{natbib}

\renewcommand{\vec}[1]{\boldsymbol{#1}}

\newacro{DM}[DM]{{Dzyaloshinskii-Moriya}}

\newcommand{\fepo}{Fe$_{3}$PO$_4$O$_3$}

\newcommand{\angstrom}{\textup{\AA}}

\begin{document}

\title{Partial Antiferromagnetic Helical Order in Single Crystal Fe$_3$PO$_4$O$_3$}
 
\author{C.L. Sarkis}
\affiliation{Department of Physics, Colorado State University, Fort Collins,CO 80523-1875}
\author{M.J. Tarne}
\affiliation{Department of Chemistry, Colorado State University, Fort Collins,CO 80523-1872}
\author{J.R. Neilson}
\affiliation{Department of Chemistry, Colorado State University, Fort Collins,CO 80523-1872}
\author{H.B. Cao}
\affiliation{Neutron Scattering Division, Oak Ridge National Laboratory, Oak Ridge, TN 37748}
\author{E. Coldren}
\affiliation{Department of Physics, Colorado State University, Fort Collins,CO 80523-1875}
\author{M.P. Gelfand}
\affiliation{Department of Physics, Colorado State University, Fort Collins,CO 80523-1875}
\author{K.A. Ross}
\affiliation{Department of Physics, Colorado State University, Fort Collins,CO 80523-1875}
\affiliation{Quantum Materials Program, Canadian Institute for Advanced Research, MaRS Centre,
West Tower 661 University Ave., Suite 505, Toronto, ON, M5G 1M1, Canada}

\date{10/18/2019}

\begin{abstract}
Magnetic frustration in \fepo\ has been shown to produce to an unusual magnetic state below T$_N = 163$~K, where incommensurate antiferromagnetic order is restricted to nanosized needle-like domains, as inferred from neutron powder diffraction. Here we show using single-crystal neutron diffraction that \fepo\ does not exhibit a preferred ordering wavevector direction in the $ab$ plane despite having a well-defined ordering wavevector length. This results in the observation of continuous rings of scattering rather than satellite Bragg peaks. The lack of a preferred incommensurate ordering wavevector direction can be understood in terms of an antiferromagnetic Heisenberg model with nearest-neighbor ($J_1$) and second-neighbor ($J_2$) interactions, which produces a quasi-degenerate manifold of ordering wavevectors. This state appears to be similar to the partially ordered phase of MnSi, but in \fepo\  arises in a frustrated antiferromagnet rather than a chiral ferromagnet.
 
\end{abstract}
\maketitle

%-----------------Introduction-------------------%

Incommensurate helical magnetic structures are of interest as potential generators of multiferroic or skyrmion phases \cite{kimura2003magnetocapacitance,rossler2006spontaneous,muhlbauer2009skyrmion}. Typically an incommensurate structure will have a well defined pitch length, $\lambda$, producing an ordering wavevector magnitude $|\vec{k}| = 2\pi/\lambda$, and a well-defined ordering wavevector direction. Both properties can be measured through neutron diffraction experiments, where Bragg peaks will be observed at satellite wavevectors $\vec{G}+\vec{k}$ around the magnetic zone centers $\vec{G}$ that correspond to the ``parent'' commensurate structure \footnote{In locally ferromagnetic incommensurate structures, those magnetic zone centers are the same as the nuclear zone centers, including $\vec{G}=(0,0,0)$, but for antiferromagnetic parent structures, the ``parent'' magnetic zone centers are not generally coincident with nuclear zone centers.}.

 Contrary to this conventional picture, there have been recent cases of interest, notably in the B20 material MnSi, in which a well-defined \emph{magnitude} for the incommensuration exists, but not a \emph{direction}. This lack of a specific direction for the ordering wavevector has been termed ``partial'' or ``unpinned'' magnetic order \cite{pfleiderer2004partial, hopkinson2009origin}. The B20 materials, e.g., MnSi, MnGe, FeGe, and Fe$_{1-x}$Co$_x$Si, host locally ferromagnetic (FM) helical phases which transform into skyrmion lattices under small applied fields \cite{muhlbauer2009skyrmion,yu2011near,adams2010skyrmion,munzer2010skyrmion,kanazawa2012possible}. The parent helical structures and field-induced skyrmion phases are long range ordered (LRO), with magnetic propagation vectors $\vec{k}$ that are well-defined in both magnitude and direction, manifesting as sharp satellite peaks in neutron diffraction experiments. The 6-fold symmetric arrangement of Bragg peaks are observed a small, constant reciprocal lattice distance away from magnetic zone centers. Due to the FM nature of the parent commensurate state, the commensurate zone centers include $\vec{G}=(0,0,0)$, so they can be most effectively studied via small-angle neutron scattering (SANS) \cite{muhlbauer2009skyrmion}. However, this well-ordered skyrmion lattice phase can be disrupted such that the six incommensurate Bragg peaks become a spherical shell of scattering due to ``partial order'': MnSi displays this type of spherical structure factor in a high pressure phase with unusual electronic transport properties \cite{pfleiderer2004partial} as well as in the correlated paramagnetic regime above its ambient pressure transition \cite{pappas2009chiral}, while Fe$_{1-x}$Co$_x$Si displays a similar shell when zero field cooled \cite{munzer2010skyrmion}. In the case of MnSi, this partial helical order has been argued to be analogous to a ``blue phase'' of chiral liquid crystals, where the directors are arranged into double-twist cylinders that are not long range ordered.  In MnSi, the analogous topological spin textures of the partially ordered phase are triple-twist cylinders, which are similar to skyrmions that are tightly packed, but not long range ordered \cite{hamann2011magnetic,tewari2006blue,henrich2011structure}. Furthermore, in MnSi this phase appears to be energetically preferred compared to a well-ordered helix \cite{pfleiderer2007magnetic}, and there is evidence that the combination of nearest neighbor FM and Dzyaloshinskii-Moriya (DM) interactions could stabilize such a state \cite{hamann2011magnetic}.  Although the case for a topologically non-trivial partial order in MnSi is compelling, it is not yet clear how general this kind of phase may be.

 %---------------------%
 \begin{figure}[!tb]
 \includegraphics[width=0.8\columnwidth]{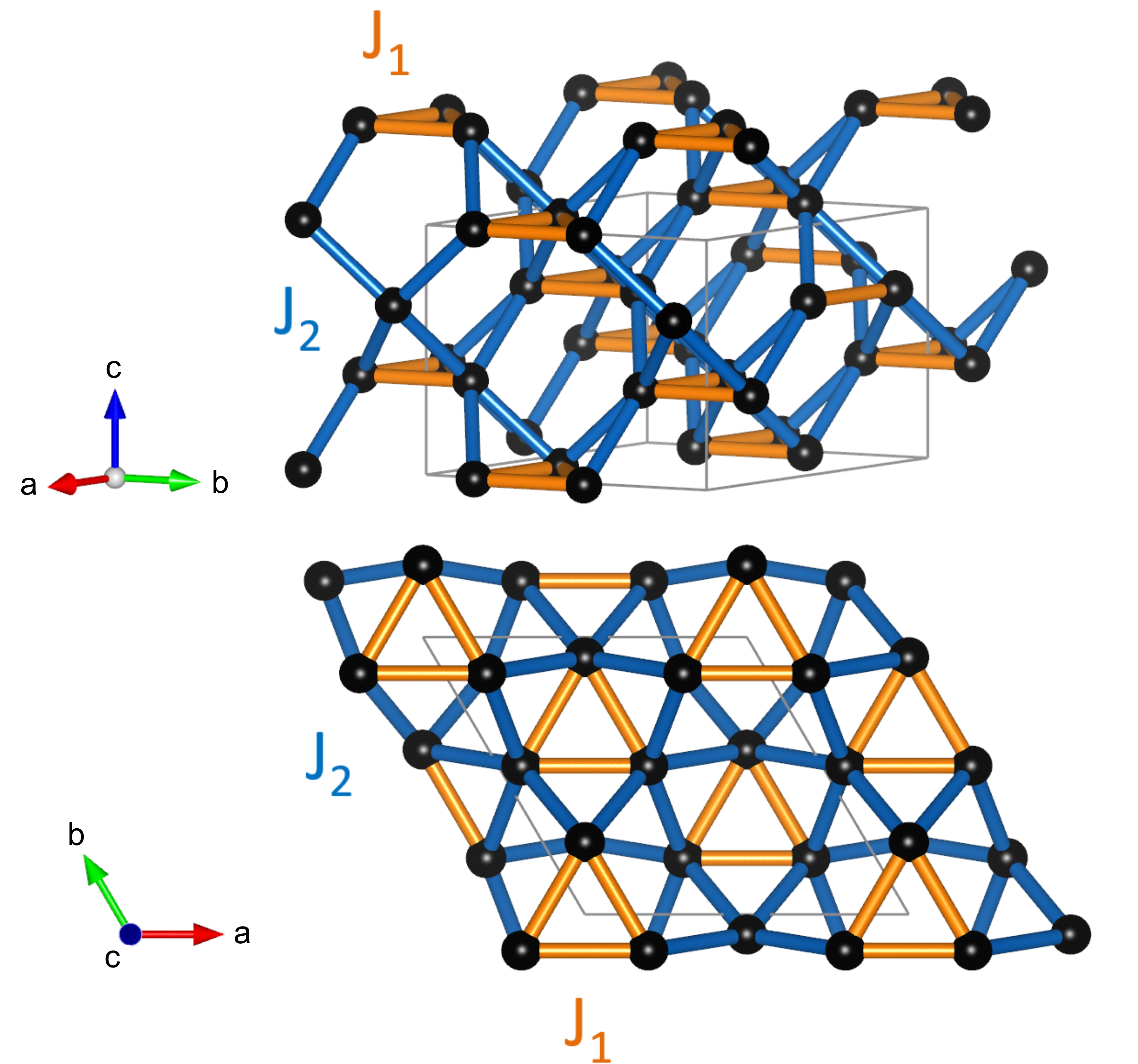}
 \caption{Magnetic sublattice formed by Fe$^{3+}$ ions in \fepo, with structural unit cell (in the hexagonal setting of $R3m$) shown as gray lines. Here $J_1$ interactions are shown in orange and $J_2$ interactions are shown in blue.}
\label{fig:structure}
 \end{figure}
 %---------------------%
 
 %---------------------------%
 \begin{figure*}
\includegraphics[width=0.8\textwidth]{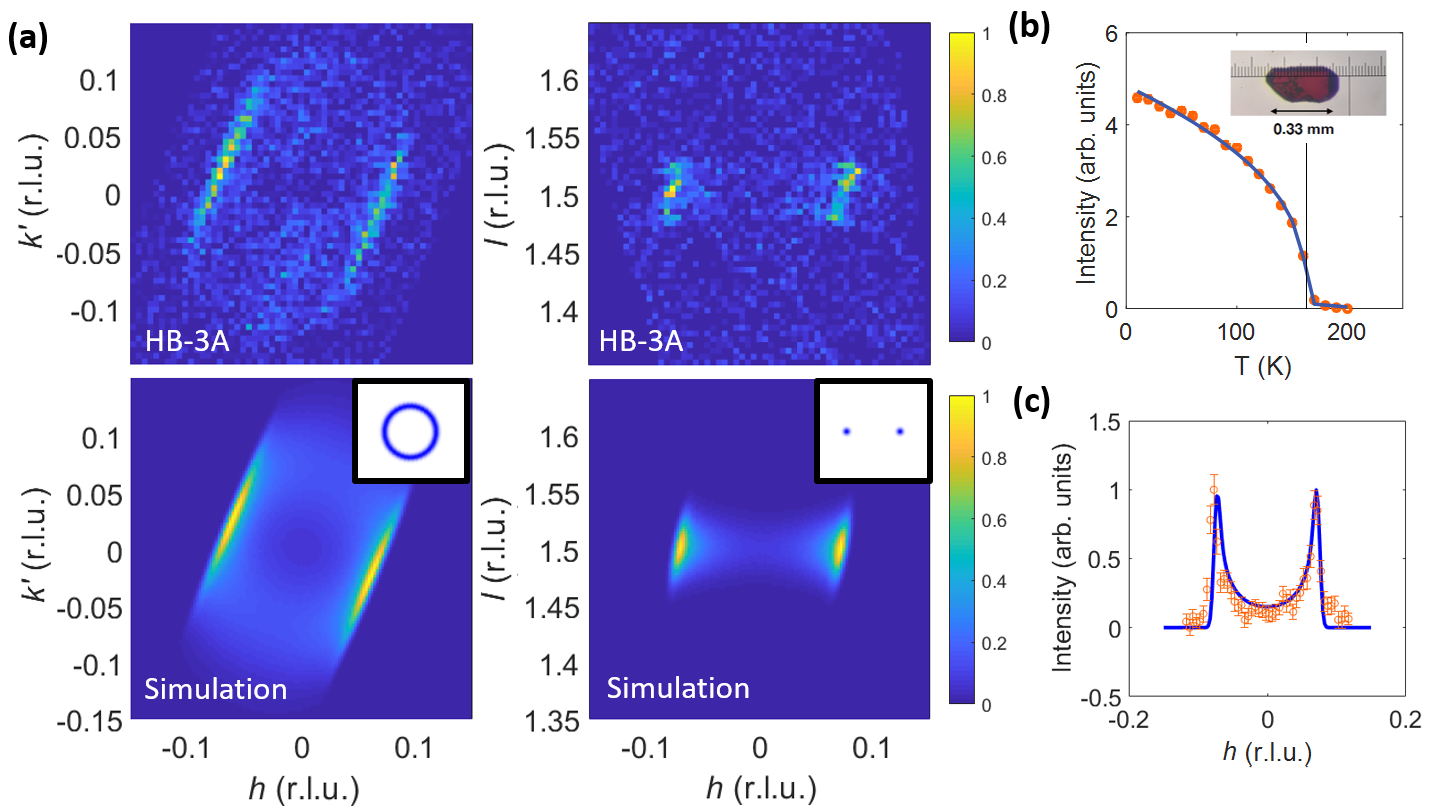}
\caption{(a) Single-crystal neutron diffraction intensity maps near $(h~k~l)$ = (0~0~1.5) at T = 4~K in two planes of reciprocal space (top). $k'$ refers to a length along the vector orthogonal to both $a^*$ and $c^*$ and normalized to be the same length as b$^*$ Calculation of the diffraction intensity expected for a resolution-limited ring of radius 98 $\angstrom$ in the $hk'$-plane and centered on $(h~k~l)$ = (0~0~1.5) (bottom). The ring is convolved with an anisotropic instrumental resolution ellipsoid that was determined via measurements of a nearby nuclear Bragg peak \cite{SI}. Insets show the non-convolved ring in each respective plane. (b) Integrated intensity of the ring as a function of temperature centered on $(h~k~l)$ = ($\bar{2}$~2~0.5). (Inset) Single crystal \fepo\ used for neutron diffraction. (c) Intensity along $a^*$ for $(h~k~l)$ = (0~0~1.5) showing agreement between data and resolution calculation.}
\label{fig:data}
\end{figure*}
%---------------------------------%
Here we report the first observation of a partially ordered helical state in a locally \emph{antiferromagnetic} (AFM) spin structure arising in the insulating magnet \fepo. The state is characterized by incommensurate \emph{rings} in the magnetic structure factor as measured by single-crystal neutron diffraction. Taken together with previous studies showing short correlation lengths in the plane of the ring, these results suggest a disordered state similar to the blue phase description of MnSi.  However, compared to MnSi, the static partial order in \fepo\ appears to be extremely stable, spanning temperatures from 163~K to 4~K (or below), with no external pressure or magnetic field required. 

The magnetic properties of \fepo\ were first investigated in the 1980's \cite{modaressi1983fe3po7,berthet1989exafs}, but study of the material has enjoyed renewed interest \cite{sobolev2018modulated,ross2015nanosized,tarne2017tuning,sarkar2017magnetic} due to recent high-resolution neutron powder diffraction (NPD) measurements. These measurements revealed unusual magnetic correlations below an antiferromagnetic transition at $T_N = 163$~K \cite{ross2015nanosized, tarne2017tuning}. \fepo\ forms a non-centrosymmetric lattice (spacegroup $R3m$, with room temperature lattice parameters of $a=8.006~\angstrom$, $c=6.863~\angstrom$ \cite{modaressi1983fe3po7} in the hexagonal setting) with magnetic Fe$^{3+}$ ions residing in a 3D network of triangular units shown in Fig. \ref{fig:structure}. Previous work on polycrystalline samples indicated strong antiferromagnetic interactions ($|\Theta_{CW}| > 1000$ K), and a frustration parameter $f = |\Theta_{CW}|/T_N \geq 6$ \cite{ross2015nanosized}. An incommensurate AFM state occurs below $T_N$ \cite{ross2015nanosized, sobolev2018modulated,modaressi1983fe3po7}, which occupies the full volume fraction of the material \cite{sarkar2017magnetic}. Analysis of M{\"o}ssbauer spectra identified an easy axis along the crystallographic $a$-axis\cite{sobolev2018modulated} with a corresponding ordered moment of 4.3 $\mu_B$. Although DM interactions are allowed by symmetry in \fepo, many properties of the state below $T_N$ seem to be captured by a simple Heisenberg model with competing nearest-neighbor ($J_1$) and second-neighbor ($J_2$) interactions \cite{ross2015nanosized,tarne2017tuning}.

Neutron diffraction on polycrystalline powders revealed magnetic peaks which are extremely broad and oddly-shaped in reciprocal space coexisting with a resolution-limited peak \cite{ross2015nanosized}. Analogous neutron powder diffraction features were also observed in some samples of the multiferroic material BiFeO$_3$ \cite{sosnowska1982spiral,sosnowska1992investigation}, but the shape of the pattern was highly sample dependent and the structure of the most ideal samples measured with high resolution neutron instrumentation were eventually understood to be a AFM magnetic cycloid structure \cite{herrero2010neutron}. In the case of \fepo, extremely high-resolution powder neutron diffraction could not resolve individual satellite peaks. This was explained by an AFM helical structure with the propagation vector residing in the $ab$ plane and a pitch length of $\lambda = 86$~\AA.  Crucially, a ``needle-like'' correlation volume was required to reproduce the data, with short range $ab$ plane correlations ($\xi_{ab}\approx 100$~\AA) and very long range correlations in the $c$ direction.  With the availability of single crystals, we can now demonstrate that the magnetic features previously identified in the powder samples do not correspond incommensurate Bragg peaks, but rather to incommensurate \emph{rings} of intensity in the $ab$ plane. Thus, a specific incommensurate wavevector direction is not selected in this material, and only partial AFM helical order exists.

%------------------------%
%----------------Experimental Sections----------%
% --------Methods and Characterization----%
Small single crystals of \fepo\ ($\sim$0.3mm on a side, see Fig. \ref{fig:data}b inset), were grown from powders of FePO$_4$ ($\geq$99$\%$ phase purity) by chemical vapor transport using ZrCl$_4$ as the transport agent (920~$^{\circ}$C at source, 800~$^{\circ}$C at sink, for 120 h) \cite{tarne2017compositional,binnewies2012chemical}. The single crystals are a transparent dark red, and are typically ~0.5$\times$0.3$\times$0.3 mm in size. The single-crystal unit cell was confirmed using a Bruker D8 Advance Quest SCXRD and Photonic Science Laue diffractometer ($a=7.998$~\AA, $c=6.854$~\AA).
%-----------Neutron Scattering-----------%
%------------------------%

Neutron diffraction on a single crystal of \fepo\ ($m \approx 0.3$ mg) was carried out on HB-3A, the four-circle neutron diffractometer at Oak Ridge National Laboratory, with an incident neutron wavelength $\lambda = 1.546~\angstrom$ from a Si-220 monochromator. The monochromator was set to doubly-focusing mode, in order to increase flux. An area detector allowed for efficient mapping of the intensity near the ``parent structure'' commensurate AFM wavevectors (e.g., (0~0~1.5) and related positions). Most of the data was taken at T = 4~K with background subtractions from T= 200~K scans. The magnetic scattering near ($\bar{2}$~2~0.5) was measured as a function of temperature in 10~K steps up to 200~K. 

Representative data at 4~K are shown in Fig. \ref{fig:data}a, along with a simulation of a resolution-convoluted ring of intensity. 
The doubly-focusing monochromator at HB-3A produces broad resolution in the vertical direction. Depending on the specific orientation of the crystal in the four-circle diffractometer for each measured ring, this broadening shows up in different directions in reciprocal space. We modeled the instrumental resolution function as an ellipsoid with Gaussian intensity profiles, whose principal axes widths were obtained from measurements of nuclear Bragg peaks (also accounting for variations in the resolution as a function of scattering angle \cite{chakoumakos2011four}). The resolution function was transformed using the four-circle diffractometer coordinate system formalism described in Busing, et al.\cite{busing1967angle} with appropriate modifications for the coordinate system at HB-3A. Further details can be found in the supplemental information \cite{SI}.

The observed diffraction patterns are consistent with a continuous ring of uniform intensity lying in the $hk$-plane. These rings appear around each parent commensurate magnetic structure zone center, and as shown in Fig. \ref{fig:data}b, they onset below the known AFM ordering transition. The rings produce two peaks in the $kl$-plane, from which we determined the radius of the ring to be $|Q|$=0.064(6)$~\angstrom^{-1}$, corresponding to a helical pitch length $\lambda = 98\pm 12~\angstrom$, which is in reasonable agreement with the pitch length determined from powder neutron diffraction (86~\AA). All five measured rings were found to be resolution limited in each dimension (see for example, Fig. \ref{fig:data} (c)) \cite{SI}. However, this does not contradict the previous determinations of correlation lengths which were made using very high-resolution powder diffraction. The best resolution at HB-3A with our setup was 0.0059 $\pm$0.00024 $\AA^{-1}$, giving a maximum resolvable correlation length of 470 $\pm$ 21 $\AA$, however the orientation of the resolution function was such that the resolution in the $hk$-plane was much coarser for typical reflections \cite{SI}. 

This measurement thus exposes a unique scattering structure factor in \fepo\ which is analogous to the spherical shells observed in MnSi. In the case of \fepo, the presence of long range \emph{commensurate} order along the $c$-axis produces a 2D ring of scattering in the $hk$-plane, rather than a spherical shell. The existence of these rings shows that \fepo\ has ``partial'' magnetic order in the $ab$ plane.
%-----------------------%
\begin{figure}[!t]
 \includegraphics[width= 0.8\columnwidth]{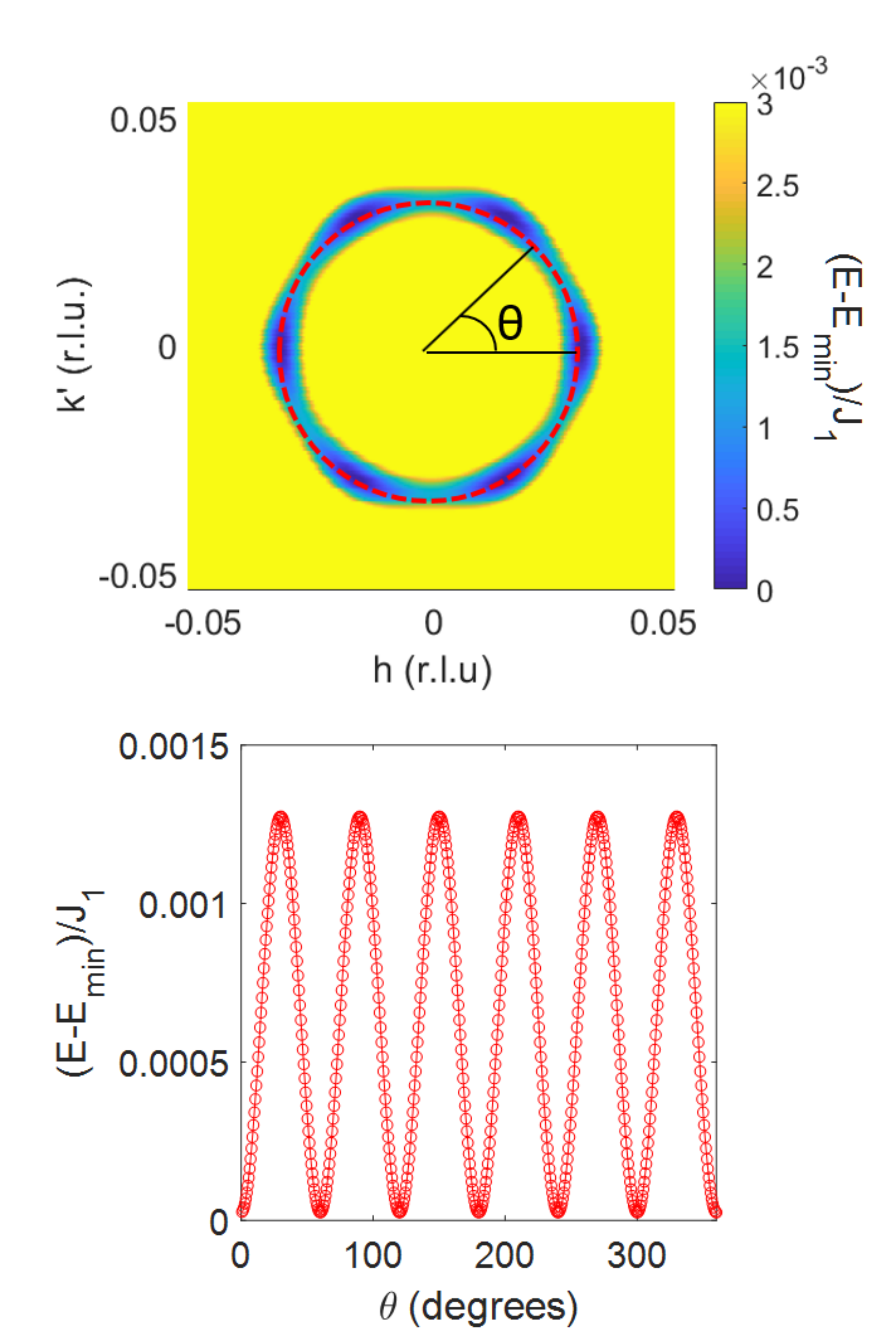}
 \caption{Luttinger-Tisza calculation of energies in the rhombehedral unit cell of Fe$_3$PO$_4$O$_3$ (but presented in the hexagonal setting) using $J_{2}/J_{1} \sim 1.9$ taken from \cite{tarne2017tuning}. Cuts within the ring show six energy minima in the $hk$-plane. Variations inside the well on the order of 10$^{-3} J_1$ correspond to energy differences of T $\sim$ 0.4~K.}
 \label{fig:LT}
\end{figure}

%%%%%%%%%%%%%%%%%%%%%%%%%%%%%%%%%%
\begin{figure*}
    \centering
    \includegraphics[width=0.65\textwidth]{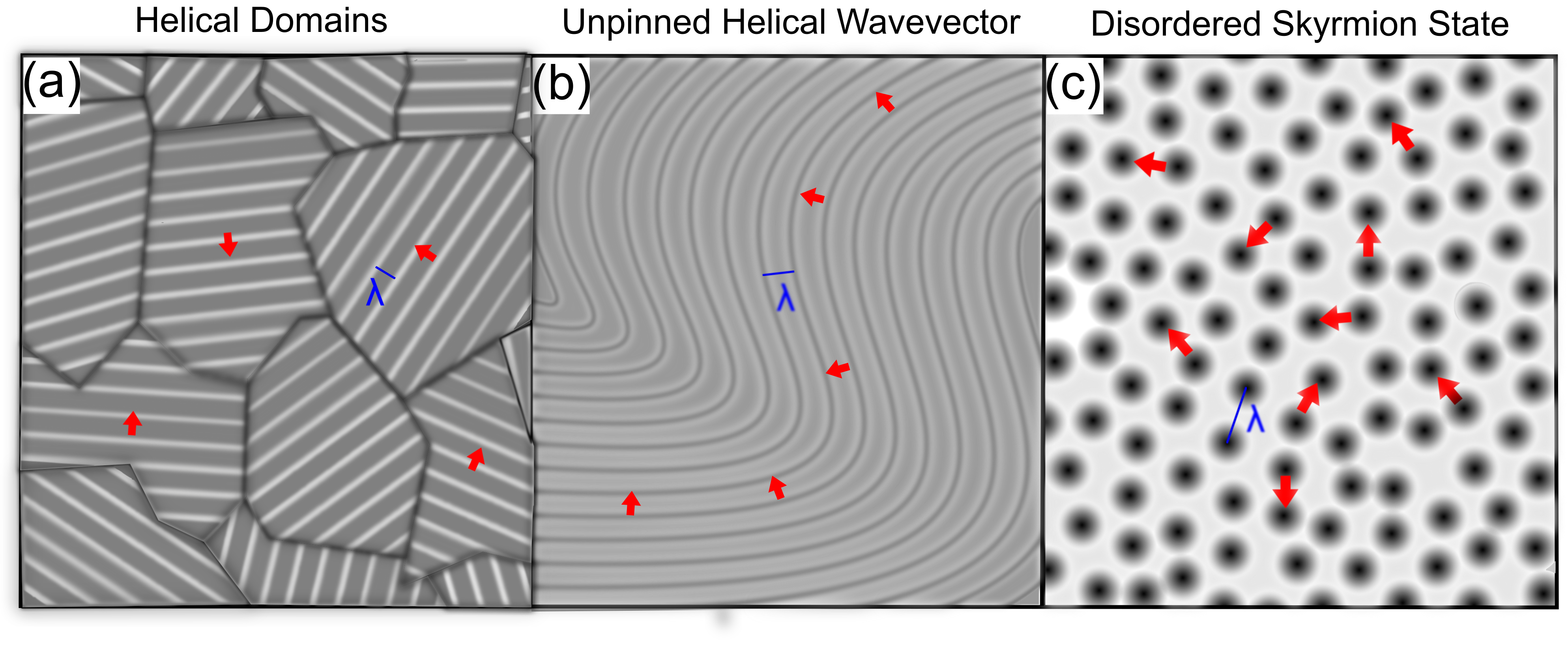}
    \caption{Possible scenarios (depicted in real space) of helical magnetic structures that can produce a circular ring in the magnetic structure factor. Lines represent the ``crests'' (zero phase angle) of an incommensurately modulated magnetic structure, with a distance of $\lambda$ (pitch length) between each line. Dots represent skyrmions. Red arrows denote incommensurate modulation wavevector directions. (a) Short range helical domains, where every domain chooses an ordering wavevector direction at random. (b) A single-domain structure with a wavevector that is unpinned from the lattice and ``meanders'' through space (similar to the blue phase in liquid crystals). (c) A disordered skyrmion state, in which skyrmions maintain an average spacing of $\lambda$ between neighbors but do not crystallize.}
    \label{fig:ring_SF}
\end{figure*}
%------------Theoretical Section----------------%
\par The lack of a preferred ordering wavevector direction in \fepo\ can be understood on the basis of the frustrated Heisenberg model used in previous studies of \fepo. We determined the ground state ordering wavevectors via the Luttinger-Tisza (LT) method \cite{luttinger1946theory,luttinger1951note,freiser1961thermal} within the frustrated AFM $J_1, J_2$ Heisenberg Hamiltonian model

\begin{equation}
 \mathcal{H} = J_{1}\sum_{n.n.} S_{i} \cdot S_{j} + J_{2}\sum_{n.n.n.} S_{i} \cdot S_{j}
\end{equation}

Where $S_{i}=\{ S_{i}^{x},S_{i}^{y},S_{i}^{z}\}$ are spins associated with site $i$ in the Fe$^{3+}$ sublattice, and $J_1$ and $J_2$ are antiferromagnetic super-exchange couplings between nearest and second neighbor pairs of spins, respectively (shown in Fig. \ref{fig:structure}). When the ``hard constraint'' of the LT method is met (i.e., all spin lengths are found to be equal), this method provides an exact determination of classical spin ground states for isotropic models \cite{lyons1960method,lyons1962classical}. 
\par Using the ratio of exchange interactions previously estimated for \fepo\ ($J_{2}/J_{1} \sim 1.9$ \cite{tarne2017tuning}) LT produces helical structures with ordering wavevectors consistent with the measured data. Examining the preferred ordering wavevectors in reciprocal space, a circular distribution of low energy configurations is obtained around the commensurate parent structure zone centers (Fig. \ref{fig:LT}). Within the model there is indeed a subtle preferred orientation within this ring (resulting in six symmetry related ordering wavevectors); however, the difference in energy of these directions compared to the rest of the circle is $\sim$  J$_1$/1000. Using an estimate of $J_1 > 319$~K based on $\theta_{CW}>1000$~K, the energy barriers between minimum around the ring is at most 0.4~K: therefore, LT predicts an effectively degenerate manifold of ordering wavevectors. 

 Given this degeneracy of the ordering wavevectors, one may imagine several scenarios for a ring-like structure factor, some of which are depicted in Fig. \ref{fig:ring_SF}. In the simplest case, short range helical domains with many topological defects (domain walls) between them could form a continuum of ordering directions with well-defined domain sizes (Fig. \ref{fig:ring_SF}a). The second case is a single-domain structure that is free to vary in direction while maintaining its pitch length (Fig. \ref{fig:ring_SF}b). The last case is a short range correlated state of topological spin textures, as in a disordered skyrmion \cite{pfleiderer2004partial} or ``blue phase'' \cite{hamann2011magnetic}, as proposed for partial order of MnSi.  In this scenario, the average distance between the topological objects would be preserved, but they would not crystallize (Fig. \ref{fig:ring_SF}c).  
 
 What remains to be understood in \fepo, which may help to distinguish between the above scenarios, is what limits the length scale of correlations within the $ab$ plane.  The influence of site dilution reported in Ref. \onlinecite{tarne2017tuning} offers a clue.  Substituting non-magnetic gallium at the iron sites of \fepo\ in powder samples of was found to cause both $\lambda$ and $\xi_{ab}$ decrease, and, intriguingly, these two lengths exhibited the same trend across the whole doping series \cite{tarne2017tuning}. This suggests that the source of the short range correlations is intrinsically tied to the microscopic origin of the incommensurate structure itself. A state like a disordered skyrmion phase would provide a natural connection between $\xi_{ab}$, the size of the coherent helical object (a skyrmion), and $\lambda$. In the case of a disordered skyrmion phase, one would expect that $\lambda$ would equal approximately twice $\xi_{ab}$, the latter being defined as the radius of the correlated region. In contrast, in the case of \fepo, it appears that $\xi_{ab}$ remains approximately equal to $\lambda$ \cite{tarne2017tuning}, suggesting a slightly different type of object in real space.  However, we note here that a quantitative determination of $\xi_{ab}$ based on the diffraction peak profile shape relies on several assumptions when the peak shape is something other than a simple Lorentzian. For instance, in Ref. \onlinecite{tarne2017tuning} the Scherrer formula was used with shape factor $K = 1$, and though this should give the right order of magnitude, this precise value of $K$, and thus the precise value of $\xi_{ab}$ is not quantitatively justified. Nevertheless, the quantitative values of $\lambda$ and the \emph{trends} in $\xi_{ab}$ reported in Ref. \onlinecite{tarne2017tuning} stand up to any choice of analysis, and strongly suggest an intrinsic connection between these length scales, as would be expected based on a disordered collection of topological objects. 

%-----------------------%

% ----------------Conclusions------------%
 
\par In summary, we have carried out detailed single-crystal neutron diffraction measurements on \fepo, which has revealed a partially ordered helical phase; i.e., one with a well-defined ordering wavevector magnitude, but not direction, down to $T=4$~K. Luttinger-Tisza calculations using the frustrated Heisenberg $J_1,J_2$ model confirm that a specific direction for helical ordering wavevector is only very weakly preferred. This result is reminiscent of the partially ordered phase in MnSi, which was argued to be an analog to the blue phase found in liquid crystals. In \fepo, this unusual state appears to also be related to a disordered AFM skyrmion-like state.  Determining the real space nature of the unusual partially ordered state in \fepo\ is thus an interesting open question.

%-------------References-------------%

\begin{acknowledgements}
	The authors acknowledge useful discussions with R. Glaum relating to the synthesis of the single crystal samples, and with D. Reznik relating to the partially ordered state of MnSi. This research used resources at the High Flux Isotope Reactor, a DOE Office of Science User Facility operated by the Oak Ridge National Laboratory.
\end{acknowledgements}

\end{document}